\documentclass[a4paper,10pt]{article}
\usepackage{graphicx}
\usepackage{epstopdf}
\usepackage{amsfonts,amsmath,amssymb}
\usepackage{hyperref}

\usepackage{epsfig,multicol}

\newcommand\fverb{\setbox\pippobox=\hbox\bgroup\verb}
\newcommand\fverbit{\egroup\item[\fbox{\unhbox\pippobox}]}

\newbox\pippobox

\begin{document}
\def\boxit#1{\vcenter{\hrule\hbox{\vrule\kern8pt
      \vbox{\kern8pt#1\kern8pt}\kern8pt\vrule}\hrule}}
\def\Boxed#1{\boxit{\hbox{$\displaystyle{#1}$}}} 
\def\sqr#1#2{{\vcenter{\vbox{\hrule height.#2pt
        \hbox{\vrule width.#2pt height#1pt \kern#1pt
          \vrule width.#2pt}
        \hrule height.#2pt}}}}
\def\square{\mathchoice\sqr34\sqr34\sqr{2.1}3\sqr{1.5}3}
\def\Square{\mathchoice\sqr67\sqr67\sqr{5.1}3\sqr{1.5}3}
\def\lambdabar{{\mathchar'26\mkern-9mu\lambda}}
\def\thrdotovervx{\buildrel\textstyle...\over v_x}
\def\thrdotovervy{\buildrel\textstyle...\over v_y}
\title{\bf Gravitomagnetism from Temporal Dimensional Reduction}

\author{{\small \ Mehran Z-Abyaneh}\footnote{me\_zahiri@sbu.ac.ir}
        {\small {}\ and
         Mehrdad Farhoudi}\footnote{m-farhoudi@sbu.ac.ir} \\
        {\small Department of Physics, Shahid Beheshti University,
         1983969411, Tehran, Iran}}
\date{\small January 8, 2026}
\maketitle
\begin{abstract}
\noindent
 We reduce
the Taub-NUT metric dimensionally to three spatial dimensions by
treating time as an extra curled dimension, and end up with the
$3$-dimensional Einstein field equations plus a corresponding
Maxwell type equations for a gravitomagnetic field, associated
with the NUT charge, which also acts as a source for the Einstein
field equations. In this approach, the Taub-NUT metric can be
envisaged as a $(1+3)$-dimensional analogue of the
$(1+4)$-dimensional metric of the Kaluza-Klein theory. Hence, in
four dimensions, it unifies gravitation and gravitomagnetism,
associated with the NUT charge, in the same footing that the
Kaluza-Klein theory unifies gravitation and electromagnetism in
five dimensions. In fact, in this way, gravity and
gravitomagnetism, associated with the NUT charge, appear as two
distinct fields that emerge from the temporal dimensional
reduction. We also introduce a relation between the
$4$-dimensional gravitational constant and the NUT charge.
\end{abstract}

\section{Introduction}
\indent

It is known that gravitation and electromagnetism can be unified
within the framework of the $(1+4)$-dimensional Kaluza-Klein (KK)
theory, which, after compactifying the extra dimension, reduces to
the Einstein equations of general relativity (GR) and the Maxwell
equations in $(1+3)$-dimensional spacetime through such a purely
geometric theory. This theory was proposed in the 1920s by two
physicists, Theodor Kaluza and Oskar
Klein~\cite{Kaluza}--\cite{Kle26b}. The main idea behind this
theory is to extend the $4$-dimensional spacetime of GR that
describes the behavior of gravity, by adding an extra dimension.
Kaluza preformed the calculations by taking into account a
cylindrical condition, that is, assuming that a fifth dimension
exists but no~physical quantities depend upon it. Another approach
to explaining the apparent $4$-dimensional nature of the universe
was introduced by Oskar Klein~\cite{Kle26a,Kle26b}, who
hypothesized that the lack of dependence on an extra dimension was
because it was compactified to a very small scale, rolled-up in a
tinny circle, explaining the fact that such an extra spacial
dimension cannot be observed. This idea, commonly known as
compactification, proposes a fifth length-like coordinate that has
a circular topology and an imperceptible
scale~\cite{dSS83}--\cite{Castillo}. This is considered as a type
of local gauge symmetry in the associated $4$-dimensional
spacetime. Hence, in KK theory, it is assumed that such an extra
dimension is responsible for the electromagnetic force, which
is~not explained by GR. In fact, this theory proposes that the
electromagnetic field is~not an independent field, but arises from
the extra dimension, similar to gravity arises from the curvature
of the four dimensions of spacetime.

The KK theory was an important precursor to modern theories of
particle physics, such as string theory and
supersymmetry~\cite{Overduin}. Although this theory has~not been
fully confirmed by experimental evidence\rlap,\footnote{Also, in
Ref.~\cite{Li2023}, it has been claimed that the $4$-dimensional
spacetime assumed in the KK theory does~not exist.}\
 it
remains an active area of research because it provides a framework
for exploring the unification of fundamental forces and the nature
of spacetime itself. Since then, it has also become one of the
essential foundations for modern theories of particle physics,
such as string theory, brane theory,
M-theory~\cite{Witten1}--\cite{Greene}, and the theories that led
to Yang-Mills theories. In fact, this theory has inspired a great
deal of research into the idea of unification of the fundamental
forces of nature. However, an ambiguous aspect of the KK theory is
the prediction of a dilaton field, which has~not yet been observed
experimentally.

On the other hand, the Taub-NUT (Newman-Tamburino-Unti)
solution~\cite{Taub,NUT} to the Einstein equations of GR has been
extensively studied in the context of gravitational physics and
mathematical relativity~\cite{Griffiths2009}--\cite{Luna}. One the
interesting aspects of this solution is the existence of the NUT
charge, which is the source of the gravitomagnetic
field\footnote{Such a gravitomagnetic field is different from the
one resulting from the weak-field and slow-motion limit of
GR~\cite{MTW}, and was experimentally confirmed in
$2011$~\cite{GprobB}.}\
 due to the
gravitomagnetic monopoles~\cite{Lynden1998}--\cite{Kol2020}, the
gravitational analogue of the magnetic monopoles~\cite{Dirac1931}.
However, a problem with the Taub-NUT solution is the string
singularity called the Misner string. One way to eliminate the
Misner string is to make the time coordinate
periodic~\cite{Misner1963}, which leads to the existence of closed
time-like curves (CTCs). CTCs are a hallmark of the Taub-NUT
solution and have profound implications for causality in GR. A CTC
is a path through spacetime that loops back on itself while
remaining time-like throughout.

However, CTCs are~not unique to the Taub-NUT metric, but appear in
several solutions of the Einstein field equations, particularly in
spacetimes with unusual topology or specific symmetry properties.
Notable examples include: The G{\"o}del solution as a rotational
cosmological model that admits CTCs~\cite{Godel1949} and violates
global causality everywhere. Also, the interior region (inside the
event horizon) of a rotating black hole, as indicated by the Kerr
and Kerr-Newman metrics~\cite{Chandrasekhar1983}, can host CTCs,
particularly near the ring singularity. These CTCs are confined to
the interior of black holes and do~not affect causality in the
external spacetime~\cite{Chandrasekhar1983}. Traversable wormholes
can create CTCs if their ends experience different time dilation
effects~\cite{MorrisThorne1988}. Certain configurations of
rotating cosmic strings can admit CTCs due to frame-dragging
effects~\cite{Deser1984}. Cylindrical distributions of rotating
dust, called the van Stockum dust, can produce CTCs, depending on
the density and angular velocity of the
dust~\cite{Vanstockum1937}. It is worthwhile to mention that
adjusting the boundary conditions at infinity is a way to avoid
CTCs~\cite{Misner1963}. Nonetheless, such an approach to eliminate
CTCs in the 4-dimensional Taub-NUT metric comes with trade-offs
such as restricting spatial coordinates~\cite{Griffiths2009}.

In this work, we investigate the Lorentzian Taub-NUT solution,
retaining its CTCs and the associated causal structure. Taking
advantage of the formal similarity of the Taub-NUT metric and the
KK metric, we show that when the NUT charge is non-zero, time can
be considered as an extra dimension that can be curled by reducing
the $(1+3)$ dimensions to $3$-spatial dimensions. The outline of
the work is as follows. In the next section, we briefly review the
KK theory as the Lagrangian dimensional reduction procedure. In
Sec.~$3$, the Taub-NUT metric and the notion of the NUT charge are
introduced. In Sec.~$4$, we explain how the unification of
gravitation and a gravitomagnetism, which originates from the NUT
charge, takes place. Finally, we summarize the results obtained in
the conclusions section. However, we include some derived
relations in an appendix.

\section{Brief Review of the Kaluza-Klein Theory }\label{sec2}
\indent

The KK theory approach to unification of electromagnetism and
gravitation involves introducing a $5$-dimensional
metric\footnote{We represent $5$-dimensional quantities with a
hat, $\hat{}$ , over them.}
\begin{equation}
(\hat{g}_{\rm AB}^{}) = \left( \begin{array}{cc}
 {g}_{\mu\nu} + \kappa^2 A_{\mu} A_{\nu}\phi^2 \; \; & \; \;
      \kappa A_{\mu}\phi^2 \\
   \kappa A_{\nu}\phi^2 \; \;                                & \; \;
      \phi^2
   \end{array} \right),
\label{5dMetric}
\end{equation}
where $\phi$ is a scalar field, the uppercase Latin indices run
from zero to four, and the lowercase Greek indices run from zero
to three. Also, the metric tensor $g_{\mu\nu}$ is exactly the
$4$-dimensional metric tensor in the Einstein theory of
gravitation, the vector $A_{\mu}$ is the electromagnetic vector
potential independent from the scalar field, $\kappa$ is a
constant that can be expressed in terms of the $4$-dimensional
Newtonian gravitational constant $G$ as
\begin{equation}\label{DefnKappa}
\kappa \equiv 4 \sqrt{\pi G}.
\end{equation}
Interestingly, a simple calculation shows that $\hat{g}=\phi^2\,
g$, where $\hat{g}$ and $g$ are each the determinant of the
corresponding metrics. The $4$-dimensional metric signature is
taken to be $(-, +, +, +)$, and we work in the natural units with
$c=1=\hbar$, in which $G$ has the dimension of the square of
length. Metric~(\ref{5dMetric}) can also be written as
\begin{equation}
d{\hat s}^2 = g_{\mu\nu}dx^\mu dx^\nu +\phi^2(dy + \kappa A_\mu
dx^\mu)^2,
\end{equation}
where $(x^\mu)=(x^0,x^1,x^2,x^3)$ are the usual four coordinates
and $y=x^4$ is the fifth coordinate.

The $5$-dimensional analogue of the Einstein-Hilbert action in
vacuum can be written as
\begin{equation}\label{act4}
\hat{ S} =\frac{1}{\hat{\kappa}^2} \int \hat{R} \sqrt{-\hat{g}}
     \, d^4 x \, dy,
\end{equation}
where $\hat{R}$ is the $5$-dimensional Ricci scalar, and
$\hat{\kappa}$ can be expressed in terms of a $5$-dimensional
counterpart gravitational constant $\hat{G}$ with the dimension of
length to the power of three in the natural units as $\hat{\kappa}
\equiv 4 \sqrt{\pi \hat{G}}$. By applying the cylinder condition,
i.e. eliminating all derivatives with respect to the fifth
coordinate, $\partial_4\, \hat{g}_{\rm AB}^{}=0$, the
$4$-dimensional effective action can be obtained through the
dimensional reduction procedure as~\cite{Overduin}
\begin{equation}
S^{\rm (KK)}\equiv \hat{ S} =\frac{1}{{\kappa^2}}\int d^4x \,
\sqrt{-g} \, \phi \, \left( R -
     \frac{1}{4} \phi^2\kappa^2 F_{\alpha\beta} F^{\alpha\beta} +
     \frac{2}{3} \, \frac{\partial^{\alpha} \phi
     \, \partial_{\alpha} \phi} {\phi^2} \right),
\label{act5}
\end{equation}
where $R$ is the Ricci scalar, and $F_{\alpha\beta} \equiv
\partial_{\alpha}A_{\beta}-\partial_{\beta}A_{\alpha}$ is the
electromagnetic field tensor. Also, we have compactified the fifth
dimension and have considered
\begin{equation}
G = \frac{\hat{G} }{ \oint dy}. \label{DefnG}
\end{equation}

Now, by varying the vacuum action \eqref{act4} with respect to the
$5$-dimensional metric $\hat{g}_{AB}$ and then performing the
dimensional reduction procedure, one obtains the corresponding
Einstein and Maxwell equations, and the modified Klein-Gordon
equation\footnote{Note that the metric signature used in this work
is different from the one used in Ref.~\cite{Overduin}.
Additionally, see the appendix for some points on deriving field
equations.}\
 \cite{ACF87,Overduin}
\begin{eqnarray}
G_{\mu\nu}  \!\!\!& = &\!\!\! \frac{\kappa^2 \phi^2}{2}
   T_{\mu\nu}^{\rm (EM)} +\frac{1}{\phi} \Big[ \nabla_{\mu}
   (\partial_{\nu} \phi) - g_{\mu\nu} \Box \phi \Big] , \nonumber \\
\nabla^{\mu} \, F_{\mu\nu} \!\!\!& = &\!\!\! -3 \, \frac{\partial^{\mu}
   \phi}{\phi} \, F_{\mu\nu}\qquad\textit{\rm and}\qquad
\Box \phi = \frac{\kappa^2 \phi^3}{4} \, F_{\mu\nu}
   F^{\mu\nu} ,
\label{4dFieldEquns}
\end{eqnarray}
where $G_{\mu\nu} = R_{\mu\nu} - R g_{\mu\nu} / 2$ is the
$4$-dimensional Einstein tensor, $T_{\mu\nu}^{\rm (EM)}\equiv
F_{\mu\alpha} F_{\nu}{}^\alpha -g_{\mu\nu} F_{\alpha\beta}
F^{\alpha\beta}/4 $ is the electromagnetic energy-momentum
tensor\rlap,\footnote{For a slowly varying scalar field, one has
$\nabla_{\mu} \, F^{\mu\nu}=0$, which is the source-free Maxwell
equations.}\
 and $\Box\equiv \nabla_{\mu}\nabla^{\mu}$.

When the scalar field $\phi$ is constant throughout spacetime,
Eqs.~(\ref{4dFieldEquns}) become
\begin{equation}
G_{\mu\nu} = \frac{\kappa^2 \phi^2}{2} T_{\mu\nu}^{\rm
(EM)},\qquad\qquad \nabla^{\mu} \, F_{\mu\nu} =
0\qquad\quad\textit{\rm and}\qquad\quad \Box \phi = 0,
\label{EinsteinMaxwell}
\end{equation}
where the first two equations are the corresponding Einstein and
the homogeneous Maxwell equations, the third equation describes a
massless Klein-Gordon scalar field known as the dilaton field. It
is worthwhile to mention that the condition $\phi$ being a
constant is valid only when $F_{\mu\nu} F^{\mu\nu} =
0$~\cite{Jor47}--\cite{Thi48}. The main achievement of the KK
theory is that starting with a source free action~(\ref{act4})
leads to Eqs.~(\ref{EinsteinMaxwell}) with a $4$-dimensional
electromagnetic source matter originating purely from the geometry
of $5$-dimensional empty spacetime.

\section{The Taub-NUT Solution}
\indent

The Taub-NUT solution is an important exact solution to the
Einstein field equations. It represents a class of spacetimes that
generalize the Schwarzschild solution by incorporating additional
parameters associated with gravitational monopoles and dipole
moments. The analogue of this solution in $5$-dimensional
spacetime is called the Gross-Perry solution~\cite{Gross1983}. The
Taub-NUT solution, which is a vacuum solution and exhibits unique
topological and geometric properties, has been expressed in the
spherical spatial coordinates as~\cite{Griffiths2009,Bardoux2014}
\begin{equation}\label{TaubNUT1}
d s^2 = -\Phi^2(r) \left(dt + 2l \cos{\theta}d{\varphi}\right)^2 +
\frac{dr^2}{\Phi^2(r)}+ (r^2 + l^2) \left(d\theta^2 +
\sin^2{\theta} d\varphi^2\right),
\end{equation}
where its metric function is obtained as
\begin{equation}\label{func}
\Phi(r) =\sqrt{ \frac{r^2 - 2 M r - l^2}{r^2 + l^2}},
\end{equation}
for when $r^2 - 2 M r - l^2>0$, which implies\footnote{However,
the range of $r$ in the spherical coordinates is $r\geq 0$.}\
 $r<
r_{-}=M-\sqrt{M^2+l^2}$ and/or $r> r_{+}=M+\sqrt{M^2+l^2}$. Here,
$M$ represents the geometric mass and $l$ the NUT charge, both of
which have the dimension of length. The NUT charge $l$
distinguishes the Taub-NUT solution from the Schwarzschild metric,
which corresponds to the case $l=0$. Such a NUT parameter
introduces a non-trivial structure in spacetime, which is often
interpreted as a gravitomagnetic monopole moment. The presence of
the term $2l \cos{\theta}d{\varphi}$ in the metric indicates a
coordinate singularity along the $z$-axis, known as the Misner
string~\cite{Misner1963}, which is a semi-infinite (quasi-regular)
singularity along the half-axis $\theta=\pi$. Such a string is
associated with non-trivial topology and is analogous to the Dirac
string in the theory of magnetic monopoles~\cite{Dirac1931}.
Unlike the Schwarzschild solution, the Taub-NUT spacetime is~not
asymptotically flat in the usual sense and exhibits more complex
asymptotic behavior due to the parameter $l$.

The main significance of the Taub-NUT solutions is the existence
of the Misner string at $\theta=0$ or $\theta=\pi$, which makes
the $1$-form $d\varphi$ ill-defined at either of these points. To
get rid of the singularity at $\theta=0$, one can shift the time
coordinate such that $t'=t+2n\varphi$, and for the point
$\theta=\pi$, one can take $t''=t-2n\varphi$, where $n$ is an
integer. Since $\varphi$ is periodic by $2\pi$, one has to
identify $t'$ and $t''$ by modulo $8\pi n$, which boils down to
imposing the periodicity to the time coordinate as $t\rightarrow
t+8\pi n$. Doing this makes the $1$-form $d\varphi$ well-defined
but leads to CTCs in the Lorentzian metric.

\section{ Time as an Extra Dimension in Taub-NUT Metric}\label{sectdefs}
\indent

In this section, we try to establish a connection between the
Taub-NUT solution and the KK framework. The main purpose is to
start from the Taub-NUT metric in a $4$-dimensional spacetime
coordinate, when $l\neq 0$, and eventually arrive at a
$3$-dimensional spatial metric plus a compactified or curled time
dimension. Actually, we assume that there is a dynamical mechanism
that drags the size of time dimension down and keeps it in a
stable way at an unobservably tiny scale. To begin, we rewrite the
$4$-dimensional Taub-NUT metric~(\ref{TaubNUT1}) in the matrix
form\footnote{We represent $3$-dimensional quantities with a
tilde, $\widetilde{}$ , over them.}
\begin{equation}
(g_{\mu\nu})  = \left( \begin{array}{cc}
-\Phi^2(r) \; \; & \; \; -2l \tilde{A}_{a} \Phi^2(r) \\
   -2l \tilde{A}_{b}\Phi^2(r)\; \; & \; \;  \tilde{ g}_{ab} -4l^2 \tilde{A}_{a}\tilde{A}_{b} \Phi^2(r)
   \end{array} \right),
\label{4dMetric}
\end{equation}
where the lowercase Latin indices run from one to three, the
solution to the $3$-dimensional vector potential is obtained as $
\tilde{\bf A}=(0,0, \cos{\theta})$, the metric signature is $( -,
+, +, +)$ provided that
$\tilde{g}_{33}-4l^2\Phi^2(r)\cos^2{\theta}>0$, and
$g=-\Phi^2(r)\,\tilde{ g}$, where $\tilde{ g}$ is the determinant
of the $3$-dimensional metric. Moreover, in the range $r< r_{-}$
and $r> r_{+}$, the solution to the $3$-dimensional metric is
\begin{equation}
(\tilde{g}_{ab})  = \left( \begin{array}{ccc}
\frac{1}{\Phi^2(r)} \; \; & \; \; 0\; \; & \; \; 0 \\
0 \; \; & \; \; r^2+l^2\; \; & \; \; 0 \\
0\; \; & \; \; 0\; \; & \; \; (r^2+l^2)\sin^2{\theta}
   \end{array} \right)
\label{3dMetric}
\end{equation}
with the signature $(+, +, +)$ in the spherical coordinates.
However, in the range $r_{-}<r<r_{+}$ including $r=l$, the scalar
field $\Phi(r)$~\eqref{func} becomes imaginary and hence the
signature of metric~(\ref{3dMetric}) changes to $( -, +, +)$.
Indeed, in these ranges, in metric~\eqref{TaubNUT1}, time
represents a spacelike coordinate and $r$ a timelike one.

Following the same steps as in Sec.~(\ref{sec2}), we proceed to
write the $4$-dimensional spacetime action in vacuum as
\begin{equation}
{S} =\frac{1}{\kappa^2} \int {R} \sqrt{-g}
     \, d^3x \, dx^0,
\label{4dAction}
\end{equation}
where $x^0$ is the Killing vector of the Taub-NUT metric and can
fulfil a compact direction, and $\kappa$ as defined in
relation~\eqref{DefnKappa}. Then, by applying the cylinder
condition, i.e. eliminating all derivatives with respect to the
$x^0$-coordinate, $\partial_0\, {g}_{\mu\nu}^{}=0$, we can achieve
the $3$-dimensional effective action from action~(\ref{4dAction})
via the dimensional reduction procedure as\footnote{The ansatz
$\Psi^2\equiv -\Phi^2$ (and hence $\Psi=i\,\Phi$) simply makes
metric~\eqref{4dMetric} similar to metric~\eqref{5dMetric} and in
turn the same for the other relations.}
\begin{equation}
\tilde {S}\equiv S =\frac{1}{\tilde {\kappa}^2}\int d^3x \,
\sqrt{\tilde {g}} \, \Phi(r) \, \left( \tilde{R} +
     \frac{1}{4} \Phi^2(r)\, 4l^2 \tilde F_{ab}  \tilde F^{ab} -
     2 \, \frac{\partial^{1} \Phi(r)
     \, \partial_{1} \Phi(r)} {\Phi^2(r)} \right),
\label{3dAction}
\end{equation}
where $\tilde{R}$ is the $3$-dimensional Ricci scalar and $\tilde{
F}_{ab} = \partial_a \tilde{A}_b-\partial_b  \tilde A_a$ is an
electromagnetic-like field tensor in three dimensions. Meanwhile,
we have analogously introduced a constant $\tilde{\kappa}$ which
can be expressed in terms of a $3$-dimensional counterpart
gravitational constant $\tilde{G}$ with the dimension of length in
the natural units as $\tilde{\kappa} \equiv 4 \sqrt{\pi
\tilde{G}}$.  Actually, we have compactified the $x^0$-coordinate
and have considered
\begin{equation}
\tilde{G} \equiv \frac{{G} }{ \oint dx^0}, \label{DefnG}
\end{equation}
where ${ \oint dx^0}$ is the result of integration along the
$x^0$-direction. The reduced action involves the gauge field
$\tilde{ A}_a$, which is a gravitomagnetic type gauge field
associated with the NUT charge, and $\Phi(r)$, which plays the
role of a scalar or a dilaton field within the context of the KK
theory.

The $3$-dimensional equations of motion from the variation of the
vacuum action (\ref{4dAction}) with respect to the $4$-dimensional
metric $g_{\mu\nu}$ and then performing the dimensional reduction
procedure, are\footnote{Also, see Ref.~\cite{Zanchin}.}
 \begin{eqnarray}
 \tilde G_{ab}  \!\!\!& = &\!\!\! -2l^2  \Phi^2(r)
    \tilde {T}_{ab}^{(\rm GM)} + \frac{1}{\Phi(r)} \left[
    \tilde{\nabla}_{a}
   (\partial_{b}  \Phi(r) ) -  \tilde{ g}_{ab} \tilde{\Box}  \Phi(r)  \right] , \nonumber \\
 \tilde{\nabla}^{a} \,  \tilde {F}_{ab} \!\!\!& = &\!\!\! -3 \, \frac{\partial^{a}
    \Phi(r) }{ \Phi(r) } \,  \tilde {F}_{ab}\qquad {\rm and}\qquad
\tilde{\Box} \Phi(r) = -l^2  \Phi^3(r) \tilde {F}_{ab}
    \tilde {F}^{ab} ,
\label{3dFieldEquns}
\end{eqnarray}
where $\tilde {G}_{ab}$ is the $3$-dimensional Einstein tensor and
$ \tilde {T}_{ab}^{(\rm GM)} \equiv \tilde {F}_{ac} \tilde
{F}_{b}{}^c -\tilde {g}_{ab}  \tilde {F}_{c d} \tilde {F}^{c d}/4
$ is a gravitomagnetic energy-momentum tensor, associated with the
NUT charge, in three spatial dimensions whose trace is non-zero,
in fact $\tilde {T}^{(\rm GM)}=\tilde {F}_{c d} \tilde {F}^{c
d}/4$. Eqs.~(\ref{3dFieldEquns}) represent the corresponding
Einstein, gravitomagnetic and Klein-Gordon equations in three
spatial dimensions, and in the appendix, we will mention some
relations and calculations in this regard. Alternatively,
utilizing some of the relations presented in the appendix, the
third equation of Eqs.~\eqref{3dFieldEquns} gives the second-order
differential equation
\begin{equation}\label{ktub2}
\Phi^2 \Phi'' + \Phi (\Phi')^2 + \frac{2\, r \Phi^2 \Phi'}{
r^2+l^2} + \frac{2 l^2 \Phi^3}{(r^2+l^2)^2} = 0,
\end{equation}
which is satisfied by the metric function~\eqref{func} as its
solution. Similarly, and also using $ \tilde{\bf A}=(0,0,
\cos{\theta})$, the $3$-dimensional metric~\eqref{3dMetric} does
satisfy the first equation in Eqs.~\eqref{3dFieldEquns}.

Our analysis indicates that when time is~not compactified, the
Taub-NUT metric unifies the gravitomagnetism and gravitation in
$(1+3)$-dimensional spacetime. On the other hand, when the time
coordinate is considered curled to a non-observable size, similar
to the fifth dimension of the KK theory, gravitation and the
gravitomagnetism, associated with the NUT charge, appear as two
distinct fields. We emphasize that such a compactification
does~not work when only the ordinary mass is involved and the NUT
charge is zero.

Furthermore, analogously comparing metric~\eqref{4dMetric} with
metric~\eqref{5dMetric}, action~\eqref{3dAction} with
action~\eqref{act5}, and Eqs.~\eqref{3dFieldEquns} with
Eqs.~\eqref{4dFieldEquns}, implies that the constant
$\tilde{\kappa}$ is likely to be related to the NUT charge $l$. In
this regard, considering their dimension (unit), we plausibly
infer such a relation to be
\begin{equation}\label{ktub}
\tilde{\kappa}^2\longleftrightarrow 2l.
\end{equation}
Accordingly, utilizing the definition $\tilde{\kappa}$ in terms of
$\tilde{G}$ into relation~(\ref{ktub}), while using
relation~\eqref{DefnG}, provides a relation between the
$4$-dimensional gravitational constant and the NUT charge as
\begin{equation}\label{ktub1}
 G\longleftrightarrow \left(\frac{\oint dx^0}{8\pi}\right)l
\end{equation}
in the natural units. By analyzing the geometric properties, such
a relation~\eqref{ktub1} between the NUT charge and the Newton
gravitational constant may be of interest.

\section{Conclusions}
\indent

The KK theory unifies gravitation and electromagnetism by adding
an extra dimension, while assuming that this extra dimension is
curled to a very small scale, such as the Planck scale. After
compactification in four dimensions, the KK metric leads to
Einstein equations and Maxwell equations of electromagnetism. On
the other hand, the Lorentzian Taub-NUT metric, as a peculiar
solution to the Einstein field equations of GR in $(1+3)$
dimensions, introduces a parameter called the NUT charge that
leads to a gravitomagnetic field, which originates from the
gravitational magnetic monopole and does~not arise in the
weak-field slow-motion limit of GR. Such a monopole exhibits a
feature called the Misner string, a singularity analogous to the
Dirac string in electromagnetism. One way to avoid the Misner
string is to make time periodic, which automatically leads to
CTCs.

In this work, we investigate the Lorentzian Taub-NUT solution,
retaining its CTCs and the associated causal structure. Then,
taking advantage of the formal similarity of the Taub-NUT metric
and the KK metric and when the NUT charge is non-zero (i.e., when
it is~not the Schwarzschild limit), we have considered time as an
extra dimension in four dimensions. After such an extra dimension
is curled, i.e. by reducing the $(1+3)$ dimensions to $(3)$
spatial dimensions, the Einstein vacuum equations in $(1+3)$
dimensions lead to the corresponding Einstein field equations in
$(3)$ spatial dimensions plus the gravitomagnetic Maxwell-type
equations related to the gravitational monopole. Such a
gravitomagnetic field also acts as a source for the
$3$-dimensional Einstein equations, similar to the KK reduction,
in which the electromagnetic field acts as a source for
$(1+3)$-dimensional Einstein equations.

In this process, we have attained gravitation and the
gravitomagnetism, associated with the NUT charge, as two distinct
fields. In fact, in this way, a new gravitomagnetic field has
emerged from the temporal dimensional reduction. Also, we have
analogously obtained a relation between the $4$-dimensional
Newtonian gravitational constant and the NUT charge.

\appendix
\section*{ Appendix}
\indent

Utilizing relations~(\ref{func}) and~(\ref{3dMetric}) and the
definition of $\tilde {F}_{ab}$, and calculating the Ricci and
Einstein tensors $ \tilde R_{ab}$ and $\tilde G_{ab}$, and the
Ricci scalar R, while using the Christoffel symbols provided in
the following, it is straightforward to show that
Eqs.~(\ref{3dFieldEquns}) hold. For example, to prove the last
equation of Eqs.~(\ref{3dFieldEquns}), utilizing the standard
relation
\begin{equation}
\tilde{\Box} \Phi = \frac{1}{\sqrt{ \mid\tilde {g}\mid}}
\partial_a \left( \sqrt{ \mid\tilde{ g}\mid} \, \tilde {g}^{ab}
\partial_b \Phi \right),
\end{equation}
its right hand-side, while using relations~(\ref{func})
and~(\ref{3dMetric}), gives
\begin{equation}
\tilde{\Box}  \Phi(r) = -\frac{2l^2 \left(\frac{ r^2 -2M r-l^2 }{
r^2+l^2 }\right)^{\frac{3}{2}}}{(r^2+l^2)^2}.
\end{equation}
On the other hand, using the definition of $\tilde {F}_{ab}$, its
non-zero components are
\begin{equation}
\tilde {F}_{23} = - \tilde {F}_{32} = -\sin{\theta},\qquad {\rm
and\ in\ turn}\qquad  \tilde {F}_ {ab} \tilde {F}^{ab} =
    2 \tilde {g}^{22} \tilde {g}^{33}  \tilde {F}_ {23}  \tilde {F}_{23} =
     \frac{ 2}{ (r^2 +l^2)^2 }.
\end{equation}
Also, from relation~(\ref{func}), we obviously get
\begin{equation}
\Phi^3(r) = \left(\frac{r^2-2M r-l^2 }{r^2+l^2
}\right)^{\frac{3}{2}},
\end{equation}
hence via these two relations, the last equation of
Eqs.~(\ref{3dFieldEquns}) is verified.

Likewise, it is easy to prove the other two equations of
Eqs.~(\ref{3dFieldEquns}). However, in this regard, using the
$3$-dimensional metric tensor~(\ref{3dMetric}) with
function~(\ref{func}), the corresponding non-zero Christoffel
symbols read
\begin{eqnarray}
 \tilde{\Gamma}^{1}{}_{ 11}\!\!\!\!\! &=&\!\!\!\!\!
    \frac{ Ml^2 -2l^2 r-Mr^2}{ (r^2 -2 M r -l^2)( r^2+l^2) },\qquad\qquad
  \tilde{\Gamma}^{1}{}_{22} =- \frac{r ( r^2-2M r -l^2)}{ r^2+l^2},
  \nonumber\\
   \tilde{\Gamma}^{1}{}_{33}\!\!\!\!\! &=&\!\!\!\!\!
   \tilde{\Gamma}^{1}{}_{22}\sin^2{\theta},\qquad\qquad\qquad\qquad\qquad\
   \tilde{\Gamma}^{2}{}_ {33} = -\cos{\theta} \sin{\theta},
   \nonumber\\
    \tilde{\Gamma}^{2}{}_{12}\!\!\!\!\! &=&\!\!\!\!\!\tilde{\Gamma}^{2}{}_{21}= \frac{r}{r^2+l^2
    }=\tilde{\Gamma}^{3}{}_{13}=\tilde{\Gamma}^{3}{}_{31},\qquad \tilde{\Gamma}^{3}{}_{23}
    = \tilde{\Gamma}^{3}{}_{32} = \cot{\theta}.
\end{eqnarray}
Hence, the corresponding $3$-dimensional non-zero components of
the Einstein tensor become
\begin{eqnarray}
 \tilde{G}_{ 11} \!\!\!\!\!&=&\!\!\!\!\! -\frac{ 2 M r^3 +3l^2 r^2+l^4}
 {(r^2 -2 M r- l^2) (r^2+l^2)^2},
 \qquad \tilde{G}_{ 22}=\frac{M r^3 +3l^2 r^2-3Ml^2 r-l^4}
 {(r^2+l^2)^2},  \cr
 \tilde{G}_{ 33}  \!\!\!\!\!&=&\!\!\!\!\! \tilde{G}_{ 22}\sin^2{\theta}.
\end{eqnarray}

It is worthwhile to mention that action~\eqref{act5} as it stands,
when $\phi$ is enclosed inside the parentheses, is actually a
Brans-Dicke action. Hence, when varying it with respect to the
$4$-dimensional metric $g_{\mu\nu}$, the scalar field $\phi$ and
the vector potential $A_\mu$, the effect of its last (kinetic)
term (i.e., $\partial^{\alpha} \phi \,
\partial_{\alpha} \phi/\phi$) appears as the additional term $(g_{\alpha\beta}\partial^{\mu}\phi
\, \partial_{\mu} \phi/2-\partial_{\alpha} \phi \,
\partial_{\beta} \phi)/\phi^2$ in the first equation of
Eqs.~\eqref{4dFieldEquns}. Whereas, such a term does~not exist
there because action~\eqref{act5} is obtained by the dimensional
reduction procedure of action~\eqref{act4}. Indeed, for the last
term inside the parentheses of action~\eqref{act5}, one can use
the relation $\partial^{\alpha} \phi \, \partial_{\alpha}
\phi/\phi^2= \Box \phi/\phi-\nabla_{\mu}(\partial^{\mu}
\phi/\phi)$. Since the second term in the right hand-side of this
relation is a complete derivative, when considered in
action~\eqref{act4}, it does~not contribute in the corresponding
field equations and can be ignored. Then, the remaining first term
in the right hand-side of this relation, when considered in
action~\eqref{act5} and $\phi$ is placed inside the parentheses,
is also a complete derivative and hence, does~not contribute in
the corresponding field equations either. Similar argument holds
for action~\eqref{3dAction} and its corresponding field
Eqs.~\eqref{3dFieldEquns}.

\section*{Acknowledgements}
\indent

The central concept and theoretical framework of this work was
conceived and developed by M. Z.-Abyaneh, who also had the primary
role in the implementation and authored the initial draft of the
manuscript. M. Farhoudi had the primary role in editing and
calculations, to which both authors contributed. Both authors
engaged in constructive theoretical discussions throughout the
project, and worked collaboratively to refine and finalize the
manuscript. M. Z.-Abyaneh also would like to thank Prof. Ingemar
Bengtsson for helpful discussions.


\end{document}